\documentclass[showpacs,preprintnumbers,amsmath,amssymb,floatfix,pre]{revtex4}
\usepackage{bm}
\usepackage{array}

\newcommand{\nabb}{{\bm{\nabla}}}
\newcommand{\gtapprox}{{\raise.3ex\hbox{$>$\kern-.75em\lower1ex\hbox{$\sim$}}}}

\usepackage[dvips]{epsfig}

\usepackage{float}

\usepackage{amsmath}
\usepackage{amsfonts}

\newcommand{\rgen}{\textrm{R}(\omega,\bar{\omega})}

\newcommand{\begineq}[1]{\begin{equation}\label{#1}}
\newcommand{\eqend}{\end{equation}}

\begin{document}
\title{Smectic blue phases: layered systems with high intrinsic curvature}

\author{B.A. DiDonna} 
\author{Randall D. Kamien}
\affiliation{Department of Physics and Astronomy, University of Pennsylvania, Philadelphia, PA 19104-6396, USA}


\begin{abstract}{
We report on a construction for smectic blue phases, which have quasi-long range smectic translational order as well as three dimensional crystalline order. Our proposed structures fill space by adding layers on top of a minimal surface, introducing either curvature or edge defects as necessary. We find that for the right range of material parameters, the favorable saddle-splay energy of these structures can stabilize them against  uniform layered structures. We also consider the nature of curvature frustration between mean curvature and saddle-splay.
}\end{abstract}
\pacs{61.30.Mp,
61.30.Jf,
02.40.-k}
\maketitle

\section{Introduction}
\label{sec:intro}

Topological defects and liquid crystal phases are hopelessly intertwined. Historically, the nematic phase derived it's name from the observation of long disclination lines. In some liquid crystal phases, such as the twist-grain-boundary (TGB) phases and blue phases, these defects are not simply artifact or nuisance, but instead act to stabilize the director field configuration. While the TGB phases have smectic order, the blue phases have purely nematic-like order created by a long range, triply periodic lattice of line defects. Two of the nematic blue phases possess cubic symmetry ($BP1$ and $BP2$) while the third ($BP3$) is thought to be an isotropic melt of double-twist cylinders \cite{spaghetti1,spaghetti2}.   
Recently, new phases of matter have been identified that possess the quasi-long range translational order of smectics~\cite{mhli} in addition to three-dimensional orientational order. These phases, dubbed the ``smectic blue phases,'' have been observed for molecules in the chiral series FH/FH/HH-$n$BTMHC, where $n$ is the aliphatic chain length. Three distinct smectic blue phases have been observed near the isotropic transition of these compounds: $BP_{smA}1$ has cubic symmetry, $BP_{sm}2$ has orthorhombic symmetry and $BP_{sm}3$ is isotropic~\cite{grelet}. The precise physical properties of these materials have been the study of intense investigation in recent years~\cite{grelet, pansubp1, pansubp2, pansubp3}.  

In general, since smectic order is incompatible with cubic symmetry, it is expected that any triply periodic crystalline structure must include smectic dislocations as well as disclinations.  However, attempts to construct smectic double-twist cylinders~\cite{kamien} and assemble them into traditional
blue phase structures~\cite{sethna} present a variety of difficulties, most notably a disagreement with
precise experimental details~\cite{pansubp2}.
In previous work~\cite{didonna} we proposed a new model for the smectic blue phases. Our construction filled space with concentric minimal surfaces wrapping a lattice of intersecting line defects. We found that
when the saddle-splay constant was large enough, these new structures were stable.  Though the new materials were chiral, our construction did not rely on macroscopic chirality as in the traditional blue phases.  Instead, the smectic compression and bending energies set the length scale of our solutions.

Here, we refine our earlier model through variations on the original construction. We find that by allowing edge dislocations our phase is stabilized for even small (negative) values of the saddle-splay constant $K_{24}$. This article is organized as follows. In Section~\ref{sec:energetics} we discuss the rotationally-invariant energetics of layered systems and derive detailed equations for the geometric frustration between curvature and uniform layer spacing. Next, in Section~\ref{sec:smallcore}, we calculate the energy and stability of likely smectic structures based on our original construction~\cite{didonna}. To supplement our analytical calculations, we present the results of simplified numerical solutions for the three dimensional smectic structure, which explore relaxation of the smectic layers away from our constructions. In Section~\ref{sec:largecore} we present a new construction for filling space based on uniform layer spacing away from a minimal surface. First we derive formulae for the evolution of curvature fields in layered space and then we employ the Weierstra\ss \ analytic representation of minimal surfaces to calculate the energies for our proposed phase. We complete our description by calculating the core energy and argue that tilt-grain boundaries form at the cores of our structures.  Numerics are also presented for this construction. Finally, in Section~\ref{sec:xray} we calculate the Fourier transform of the smectic density and compare it with experimental X-ray results.  In Appendix A we derive curvature evolution equations in curved space and in Appendix B we review the Weirstra\ss \ representation for completeness.

\section{Energetics and Construction}
\label{sec:energetics}

The similarity between the crystal structures and phase diagrams of the nematic and smectic blue phases suggests that, like the nematic blue phases, the smectic phases are stabilized through saddle-splay. The key to our construction is the observation that saddle-splay and Gaussian curvature are identical~\cite{rmp} for layered systems with uniform spacing. The saddle-splay energy of a unit director field ${\bf N}$ is~\cite{sethna}
\begin{equation}
F_{\rm SS} = K_{24} \int d^3 x \,\nabb \cdot \left[ \left( {\bf N} \cdot \nabb \right) {\bf N} - {\bf N} \left( \nabb\cdot {\bf N} \right) \right],
\label{eq:ssd}
\end{equation}
where $K_{24}$ is a Frank constant. In a layered system, we can rewrite the expression for $F_{\rm SS}$ in a more useful form by employing a local coordinate system where one direction is parallel to the local layer normal, ${\bf N}$. This frame is appropriate in the limit where the nematic director and the layer normal are locked. Then Eq.~(\ref{eq:ssd}) becomes
\begin{equation}\label{eq:ssad}
F_{\rm SS} = -2K_{24} \int \!dn\!\int dxdy\sqrt{g_n(x,y)}\,a_n(x,y)K_n(x,y)
\end{equation}
where $n$ is the Lagrangian coordinate which labels the layers, $a_n(x,y)$ is the local layer spacing at $(x,y)$, $K_n$ is the Gaussian curvature of the $n^{\rm th}$ surface, and $g_n$ is the determinant of the two-dimensional, induced, surface metric. In the special case that $a_n(x,y)$ is constant, the integral becomes purely topological:  a consequence of the Gauss-Bonnet theorem~\cite{rmp} is that for a surface of genus $g$ the integrated Gaussian curvature is $4\pi(1-g)$. Since $g > 0$ for any infinite surface~\cite{foot1}, $F_{\rm SS}$ is large and negative when $K_{24}<0$ and the unit cells contain surfaces with large genus, {\sl i.e.} many handles and holes. 
Note that here, the saddle-splay is a measure of the layer normals and not the nematic director.  When the nematic director follows the layer normal these are, of course, equivalent.  As discussed in \cite{didonna}, in type-II smectics it is possible for the saddle-splay of the director field to differ in its precise numerical value from the saddle-splay in the layers.

In addition to the saddle-splay, we must include the rotationally invariant bulk free energy:
\begin{equation}
\label{eq:smfe}
F_{Sm} = \int d^3x \, \left\{\frac{B}{4}\left[\left(\nabb\Phi\right)^2-1\right]^2 + 2 K_1 H^2\right\}
\end{equation}
where the smectic density is $\rho\propto\cos\left(2\pi\Phi/a_0\right)$, $\Phi(x,y,z)$ is a phase field, $a_0$ is the layer spacing, $B$ is the compression modulus, $K_1$ is the bend modulus, and $H=\frac{1}{2}\nabb\cdot{\bf N}$ is the mean curvature of the layers.   If the two principle curvatures are $\kappa_1$ and $\kappa_2$, $K_G=\kappa_1\kappa_2$ and $H=(\kappa_1+\kappa_2)/2$~\cite{rmp}.

Together, Eqs.~(\ref{eq:ssad}) and (\ref{eq:smfe}) favor configurations with uniform layer spacing, low mean curvature and high Gaussian curvature. However,  there is an unavoidable geometric frustration between these three terms.   
When a surface is displaced along its normal by $\delta a$, the changes in the metric and curvature tensors are~\cite{guven}:
\begin{eqnarray}
\delta g_{ij} &= &2 \kappa_{ij} \delta a \nonumber \\
\delta \kappa_{ij}& =& - \nabla_i \nabla_j \delta a + \kappa_{ik} \kappa^k_{\ j} \delta a
\label{eq:deltag}
\end{eqnarray}
It is straightforward to find the variation in the mean and Gaussian curvatures:
\begin{eqnarray}
\delta H &=&  \left[K_G - 2H^2\right] \delta a - \frac{1}{2} g^{ij} \nabla_i \nabla_j 
\delta a \nonumber \\
\delta K_G &=& -2K_G H \delta a - 
\tilde{\kappa}^{ij}  \nabla_i \nabla_j \delta a.
\label{eq:deltacurv}
\end{eqnarray}
where $\nabla_i$ is the covariant derivative on the surface and $\tilde{\kappa}^{ij}\equiv\epsilon^{ik} \epsilon^{jl} \kappa_{kl}/{\vert g \vert}$, where $\epsilon^{ij}$ is the completely antisymmetric tensor. 

If we impose smectic order, then $a$ may also be interpreted as the normal distance between adjacent layers. When the spacing is uniform, ${\bf N} \cdot \nabla \Phi$ is constant and $\nabla_i \delta a = 0$. In this case, the first result in Eq.~(\ref{eq:deltacurv}) implies $\partial H / \partial a = 0$ if and only if the principal curvatures $\kappa_i$ satisfy
\begin{equation}
\kappa_1^2 + \kappa_2^2  = 0 
\end{equation}
{\sl i.e.} $\kappa_1=\kappa_2=0$.  Thus if $K_G\ne 0$ and the layer spacing is uniform, then $\partial H / \partial a \ne 0$; Gaussian curvature leads to mean curvature.
 
For uniform spacing the evolution equations become particularly simple. The variation $\delta a$ is constant and the equations can be integrated:
\begin{gather}
H(a) = \frac{H+a K_G}{1+2aH+a^2K_G} \nonumber \\
K_G(a) =  \frac{K_G}{1+2aH+a^2K_G},
\label{eq:curvvsa}
\end{gather}
Alternatively, this evolution follows from the observation that if the principle radii of curvature are $R_1=\kappa_1^{-1}$ and $R_2=\kappa_2^{-1}$ for
one surface then the radii for the surface displaced by $a$ along the local normal are
$R_i(a)=R_i+a$. 
Thus, in layered systems, Gaussian curvature in one region implies mean curvature in another. Furthermore, as we continue to develop the initial layer, there will be a curvature singularity at a distance $a=\left(-H\pm\sqrt{H^2-K_G}\right)/K_G$ normal to the original surface, {\sl i.e.} at one of the radii of curvature.  

\begin{figure}
\center

\epsfig{file= 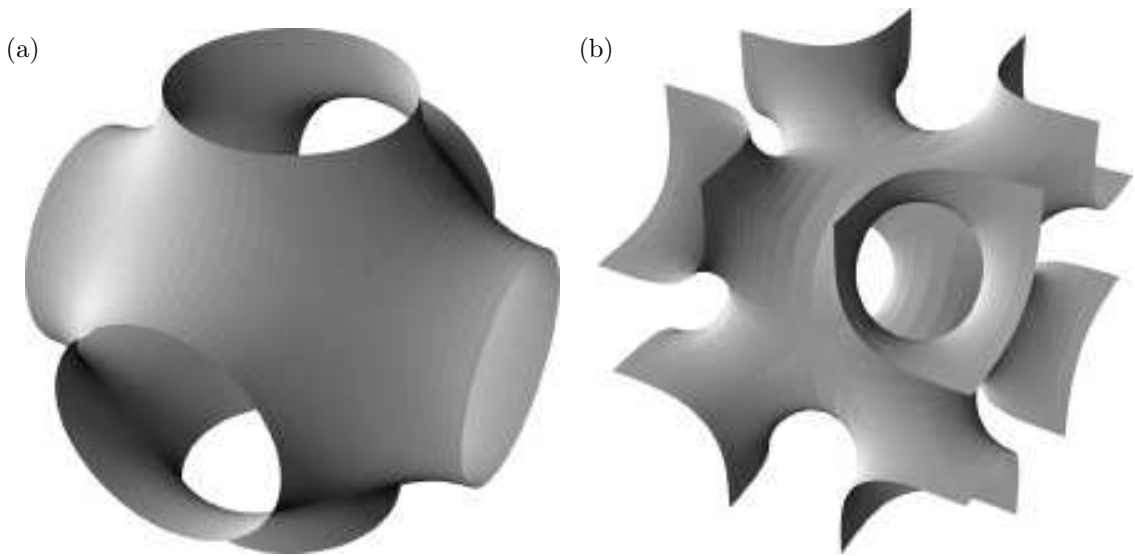}

\caption{Minimal surface repeat units: (a) P~surface, (b) I-Wp surface.}
\label{fig:surfaces}
\end{figure}

With these geometric constraints in mind, we construct a solution which strikes a promising balance between mean and Gaussian curvatures. To do this, we build the smectic order by adding layers on top of a triply-periodic minimal surface. By starting with a minimal surface, we bias our structures to have a low total mean curvature energy. In~\cite{didonna}, we based our construction on the Schwartz P~surface, pictured in Fig.~\ref{fig:surfaces}(a). In this paper, we present calculations for both the P~surface and Schoen's I-Wp~surface, pictured in Fig.~\ref{fig:surfaces}(b), a surface with a larger genus and thus a lower saddle-splay energy. The P~surface is an archetypal triply-periodic minimal surface and will provide a basis of comparison with earlier results, while the I-Wp~surface shares the symmetries of experimental $Sm_{BP}$ systems~\cite{pansubp2}; the latter will also prove to be more stable than the P~surface.  We will consider two variations of this construction: in the first we will allow the layer spacing to vary, while in the second we do not. In the former we find that
the cores are composed of topological line defects, while in the latter we find that domain
walls form in the relatively larger cores.

\section{Small core model: Self similar layering}
\label{sec:smallcore}
\subsection{Geometric Construction}
\begin{figure}
\center

\epsfig{file=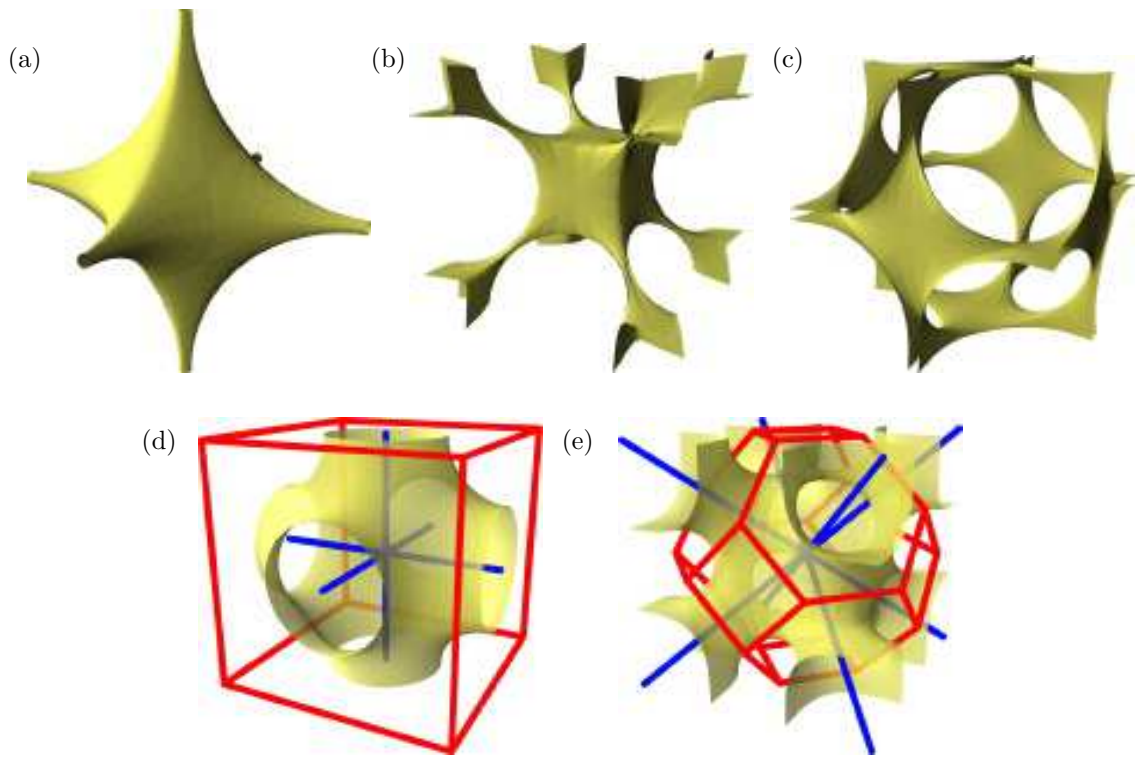}




\caption{In (a) we show the P~surface translated inward until self intersection. In (b) and (c) The I-Wp~surface is translated inward and outward, respectively. Figures (d) and (e) indicate the line defect structure for our construction, based on skeletal graphs.}
\label{fig:skeleton}
\end{figure}

In this section we consider space-filling, layered structures in which each layer is continuous and every layer has the same global topology.  We begin with a minimal surface and devise an explicit construction for filling the region away from that surface. We note that our bulk phase must contain line defects, since at some point the curvature will diverge, as is implied by Eqs.~(\ref{eq:curvvsa}). By design, the layers immediately surrounding these line defects will have the same topology as the original surface. Therefore, the next logical step in our construction is to guess the optimal line defect structure. 

Besides selecting the correct topology, we should choose a defect complexion that minimizes the compression energy in the region between the minimal surface and the defects: it should, as much as possible, be equidistant from the minimal surface at each point. Thus, an {\sl ansatz} for the optimal structure is generated by uniformly translating the minimal surface along its normal until it self-intersects, as shown in Fig.~\ref{fig:skeleton}.  The line defects will be disclinations of charge $+1$ or $+1/2$. In the case of charge $+1$ disclinations the layers immediately surrounding them are tight cylinders. From Fig.~\ref{fig:skeleton}(a), we see that the defects in the P~surface are all $+1$. From Fig.~\ref{fig:skeleton}(b) and (c), we see that the defects inside the I-Wp surface should have charge $+1$, but that outside, the translated layer approaches itself in a plane rather than on a line and so the disclination charge is $+1/2$.  Fig.~\ref{fig:skeleton}(d) and (e) show our proposed defect structures. For the defect structure outside the I-Wp surface, the location of the intersection of defect lines is a variable chosen to minimize the total compression energy. It should be noted that in the case where all defects happen to be charge $+1$, our requirements of topology and equal spacing are exactly those best satisfied by the skeletal graph of the minimal surface~\cite{visuals}. We take the core region around the defect lines to have a radius of order $a_0$, the smectic layer spacing.  We model the core as a melted region of smectic with a free energy arising from the condensation energy.

Finally, we fill the region between minimal surface and defects with continuous layers. We desire to have as little mean curvature as possible in this region, so we fill the region with dilated copies of the original minimal surface. In order to do this, we must cut the continuous minimal surface into smaller patches: we cut the surface along lines where one surface tangent is parallel to the line defect structure, and dilate it so that it shrinks onto the vertices where defect lines meet. The separated minimal patches are connected together by cylindrical patches parallel  to the line defects. When there are disclinations, we must also add some flat patches to complete the surfaces. Thus the surface tangents are continuous everywhere, and the intermediary structures transform smoothly from the minimal surface on the outside to a set of cylinders surrounding line defects on the inside. This construction fills space completely, at the expense of uniform spacing.  Moreover, we have filled a large part of space with minimal surfaces and so the curvature energy is likely to be as small as possible, given the necessary curvature defects.

\begin{figure}
\center

\epsfig{file=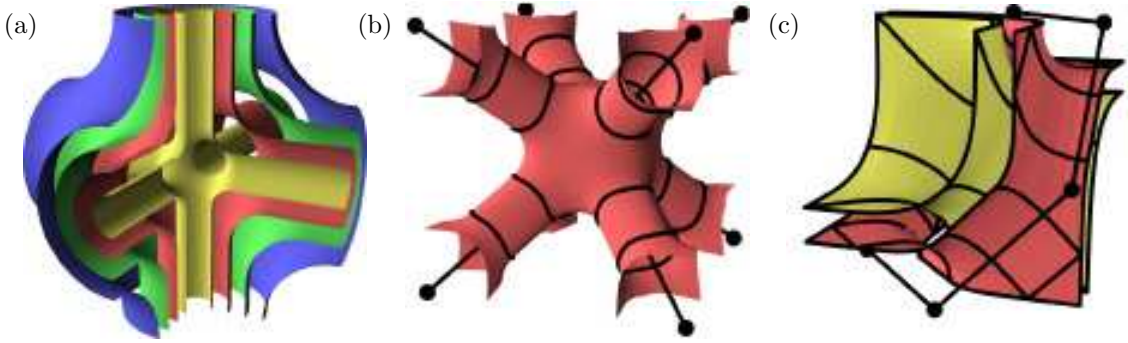}

\caption{Small core construction: (a) Partial layers inside the P~surface, (b) One layer inside the I-Wp~surface, and (c) Partial layer around one octant outside the I-Wp~surface. Each color corresponds to a different connected layer. In (b) and (c), the boundaries between minimal surface patches, cylinders and planes are drawn in heavy black. The defect networks and vertex points are also shown.}
\label{fig:smallcore}
\end{figure}

Fig.~\ref{fig:smallcore} shows how this construction works for the P and I-Wp surfaces. For the P surface it is particularly simple; its construction is show in Fig.~\ref{fig:smallcore}(a). The line defects are along the edges of the unit cell and through the center parallel to the edges. The inside of the P~surface is filled with dilated copies of the P~surface repeat unit, cut at its intersection with the walls of the unit cell. The edges of the dilated P~surface copies are connected to neighboring cells with cylinders. The outside of the P~surface is a translated copy of the inside. The I-Wp surface is more complicated, requiring different constructions for outside and inside as well as having both $+1$ and $+1/2$ defects. Fig.~\ref{fig:smallcore}(b) illustrates the construction inside the I-Wp~surface: the minimal surface is divided at cuts that can be smoothly attached to cylinders, then dilated towards either the center or corners of the unit cell.  Fig.~\ref{fig:smallcore}(c) shows the construction outside the I-Wp surface. The surface is cut along its mirror symmetry planes, $x = y$, $x=-y$, $y=z$ etc. and $x=0$, $y=0$ and $z=0$. The resulting patches are dilated around the proper vertices in Fig.~\ref{fig:skeleton}(e). The edges of the patches are connected together with cylinders parallel to the $+1/2$ edge defects. Finally, the free edges of the cylinder are connected together into planes.

\begin{table}
\center

\renewcommand{\arraystretch}{1.5}

\begin{tabular}{| c | c | c | c | c | c | c |}\hline
surface & genus & $F_B / (K_1 L)$ & $F_{SS} / (\vert K_{24} \vert L)$  & $F_{core} / \varepsilon$ & 
$F_C / (B L^3)$\\ \hline 
P & 3  & $6 \pi \left[\log (L/4\rho_c) + \left(4\rho_c/L-1\right)\right]$ & $-12\pi \left(1-4\rho_c/L\right)$ &
 $6(L - 4 \rho_c)+(64/\pi) \ \rho_c$ &  $4.32 \times 10^{-2}$ \\
I-Wp & 7 & $31.7 \left[\log (L/6.3 \rho_c) + \left(6.3 \rho_c/L-1\right)\right]$ & $-68.9\left(1-6.3 \rho_c/L\right)$ & $9.05(L - 6.1 \rho_c))+72.2 \ \rho_c $& $4.85 \times 10^{-2}$ \\ \hline 
\end{tabular}

\caption{Smectic energy per unit cell for P and I-Wp surface constructions. The first column gives the topological genus per unit cell of the structure. $F_B$ is the total curvature energy, $F_{SS}$ is the saddle-splay, $F_{core}$ is energy of the defect cores, and $F_C$ is the total compression energy.}
\label{tab:smallcore}
\end{table}

Most of the energetics of this structure are simple to calculate; exact values are tabulated in Table~\ref{tab:smallcore}. It is instructive to derive approximate expressions for these energies so that we may quickly assess the feasibility of a given minimal surface as a starting point for our construction.

The total curvature energy comes from the contribution of the cylindrical connecting patches. It is roughly
\begin{eqnarray}
F_B &\approx&  \frac{1}{2} K_1 \int \frac{l (R-r)}{Rr} \, dr \, d\theta \nonumber \\
&\approx& \pi K_1 \left[\sum_{lines}l_i c^2_i \right] \left(\log \left(\frac{\left<R\right>}{\rho_c}\right) + \left( \frac{\rho_c}{\langle\,R\,\rangle} - 1\right)\right),
\label{eq:fb}
\end{eqnarray}
where the sum is over defect lines, $l_i$ is the length of line $i$, $c_i $ is the charge of the defect, $\rho_c$ is the core diameter of the line defect, and $\langle\,R\,\rangle$ is the average radius of cylindrical regions. The factor of $l (R-r)/R$ above arises from the dependence of the cylinder length integration on the radius $r$.
The core energy, which is simply a condensation energy density, is approximately:\begin{equation}
F_{core} \approx \varepsilon \left[\sum_{lines} l_i  c^2_i \right] \left(1 - \frac{\rho_c}{\left<R\right>}\right) + 
\varepsilon \frac{L^3}{\left<R\right>^3} \frac{\rho_c}{\pi},
\label{eq:fcore}
\end{equation}
where $\varepsilon$ is a line tension. The first term is the energy of the defect lines, the second is the energy of the central, melted region around vertices in the defect network.

Since the saddle-splay is a surface term, we calculate it by integrating the strength of the nematic defects along the total length of defect core.  Because the saddle-splay is a total divergence, the volume integral becomes a surface integral around the defect:\begin{eqnarray}
F_{SS} &=& 2 K_{24} \int H \, dS \nonumber \\
&\approx& 2 \pi K_{24}\left[\sum_{lines}l_i c_i \right] \left(1-\frac{\rho_c}{\left<R\right>}\right).
\label{eq:fss}
\end{eqnarray}
Comparison to Eq.~(\ref{eq:fb}) shows that these smectic structures will favor charge $+1/2$ structures when possible. Also, for all constructions which only utilize $+1$ defects, the ratio of saddle-splay to curvature energy will be approximately $2 K_{24}/K_1$.

The only energy not explicitly determined by our construction is the compression energy. Though the layering is largely constrained, there is still one free variable which gives the relative dilation of consecutive surfaces. We minimize the compression energy with respect to the relative smectic layer spacing as a function of position. In the following we show how we can take advantage of the dilational symmetry of our construction to greatly simplify the exact calculation of optimal compression energy. 

First we consider the compression energy in a subregion which is filled with self-similar, radially dilated patches of minimal surface. Each successive layer in this region is a smaller version of the last, and all are centered on a common origin. Thus, if the radial coordinates of the outermost surface of this region are specified by a function $r_0(\theta,\phi)$, then the radial coordinates of any interior surface are given by $r = \zeta r_0(\theta,\phi)$, where $\zeta\in[0,1]$ parameterizes the layers.  Furthermore, every point within this region has a unique value of $\theta, \phi,$ and $\zeta$, and conversely the values of $\theta, \phi,$ and $\zeta$ uniquely specify the position of any point in the region, so we can use these three variables as a coordinate frame on the patch. In these coordinates, each layer is a surface of constant $\zeta$.

In the continuum description of the smectic, the layers are surfaces of constant $\Phi$.  Since $\Phi$ is constant on each smectic layer, which is, in turn, a layer of constant $\zeta$, it follows that $\Phi = \Phi (\zeta)$ is purely a function of $\zeta$. Our goal is to find the form of $\Phi (\zeta)$ which minimizes the compression energy. This is non-trivial, since the compression energy goes as $\nabla \Phi$, which is not purely a function of $\zeta$. However, we can find a combination of $\vert \nabla \Phi \vert $ times a function of angle which together is constant over each smectic layer. First we observe that
\begin{eqnarray}
\Phi(\zeta) &=& \Phi\left(\frac{r}{r_0\left(\theta, \phi \right)}\right) \nonumber \\
&\Rightarrow&\frac{\partial}{\partial\zeta}\Phi(\zeta) = r_0\left(\theta, \phi \right) \frac{\partial}{\partial r}\Phi\left(\frac{r}{r_0\left(\theta, \phi \right)}\right)\end{eqnarray}
Then we note that for level surfaces of $\Phi$ the field of unit normals is ${\bf N} = \nabb \Phi / \vert \nabb \Phi \vert$ to write:
\begin{equation}
\left({\bf r} \cdot {\bf N}\right) \vert \nabb \Phi \vert = r \frac{\partial}{\partial r} \Phi\left(\frac{r}{r_0\left(\theta, \phi \right)}\right)  
 =  \frac{r}{r_0\left(\theta, \phi \right)}\frac{\partial}{\partial \zeta} \Phi(\zeta)
  =  \zeta  \frac{\partial}{\partial \zeta} \Phi(\zeta)
\end{equation}
This shows that the combination $\left({\bf r} \cdot {\bf N}\right) \vert \nabb \Phi \vert$ depends only on $\zeta$, and so is constant over each smectic layer. We can therefore write
\begin{equation}
\left(\nabb \Phi\right)^2 = p( \theta, \phi ) \Delta ( \zeta ),
\label{eq:deltadefine}
\end{equation}
with
\begin{eqnarray}
p( \theta, \phi ) &=&\left[ \frac{{\bf r}_0 (\theta_0, \phi_0) \cdot {\bf N} (\theta_0, \phi_0)}
{{\bf r}_0 (\theta, \phi) \cdot {\bf N} (\theta, \phi)}\right]^2\\
\Delta ( \zeta ) &=& A_N \left[\nabb\Phi\right]^2{\Big\vert}_{\zeta,\theta_0, \phi_0},
\label{eq:delzetadef}
\end{eqnarray}
where $\left( \theta_0, \phi_0 \right)$ is a reference direction and $A_N$ is a normalization constant which adjusts for differences between reference directions on different patches. The absolute value of  $A_N$ is unimportant, but its relative value on different surface patches must be chosen for a consistent definition of $\Delta ( \zeta )$ across the entire smectic surface. Thus we have separated the angular dependence out of the volume integral of $\nabla \Phi$. The result is
\begin{equation}
F_{C \ patch} = \frac{1}{2} B
\int d \zeta  \left[ \zeta^2\left( I_{0} - 2 I_{1} \Delta (\zeta)  +  I_{2} \Delta^2 (\zeta)\right)\right],
\label{eq:fcpcell}
\end{equation}
where the $I_N$ are moments of the minimal surface shape,
\begin{equation}
\nonumber I_{N} = \int d \Omega \ r^3_0(\theta, \phi) p^N (\theta, \phi)
\end{equation}
Simple consideration shows that a similar separation occurs in regions with cylindrical symmetry, in which case $r=\zeta r_0(\phi)$ and $\vert \nabb \Phi \vert$ is independent of the local $z$ coordinate. The equations are similar to the above. 
Numerical values were calculated with the aid of the the Surface Evolver software package~\cite{Brakke92}.
We minimize the total $F_C$ by varying with respect to $\Delta(\zeta)$. The results are tabulated in Table~\ref{tab:smallcore}. The total compression energy had only a very weak dependence on the core size $a_0$ for
$a_0 \ll L$.

Calculations of the compression energy depend greatly on the particulars of the surface shape, so we cannot give an accurate approximate expression for $F_C$. In general, however, the total compression energy should increase rapidly with genus. Since all layers in this construction have the same topology, there must be just as many layers in the cylindrical regions, of radius $\left<R\right>$, as there are in the minimal surface patch regions, which have typical radius of order $L/4$.  As the genus increases, the number of cylindrical handles increases, as will the difference between $\left<R\right>$ and $L/4$. This, in turn, will increase the overall compression energy.

For our proposed smectic structure to be stable against the uniform flat phase, the positive energy contributions from $F_C$, $F_B$, and $F_{\rm core}$ must be compensated by a large negative saddle-splay energy. Since there is no chirality in our construction the unit cell length can only arise as a result of the different scalings of these energies. We found that the scaling competition between the core and saddle-splay energies was capable of generating a length scale of $L=50a_0$, giving a preferred cell size for the P and I-Wp surfaces  on the order of:
\begin{eqnarray}
N_{\hbox{\small\sl P}}&=&\frac{ L }{a_0} \approx\frac{12\pi \vert K_{24} \vert + \left[16/\pi - 6\right]\varepsilon}{2\pi \vert K_{24} \vert -   \varepsilon}. \\
N_{\hbox{\small\sl I-Wp}}&=&\frac{ L }{a_0} \approx \frac{72 \vert K_{24} \vert - 2.8 \varepsilon}{7.6 \vert K_{24} \vert -   \varepsilon}.
\end{eqnarray}

\begin{figure}
\center

\epsfig{file= 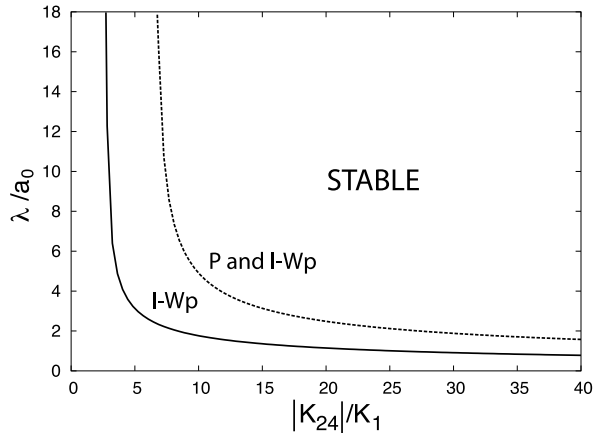,width=3in}

\caption{Stability diagram for unit cells with $L=50a_0$. Our constructions based on P and I-Wp surfaces can be stabilized against flat layering in the regions of parameter space to the upper right of the respectively labeled lines. The quantity $\lambda \equiv \sqrt{K_1/B}$ is the penetration length.}
\label{fig:tubestable}
\end{figure}
We compare the total energy of the minimal surface smectics against that of uniform smectic configuration. Fig.~\ref{fig:tubestable} shows the exact stability diagram of both the P and I-Wp surfaces, assuming a crystal unit cell length $L=50a_0$ to make contact with experiment. There are stable solutions of the P~surface smectic for values of $\vert K_{24}\vert/K_1 > 6.5$ and for the I-Wp~surface smectic for  values of $\vert K_{24}\vert/K_1 > 2.7$.

\subsection{Numerical Minimization}
\label{sec:simulation}

In order to study elastic relaxation away from our simplified construction, we performed a numerical minimization of the smectic energetics of Eq.~(\ref{eq:ssd})~and~(\ref{eq:smfe}) on a 3-dimensional grid. The smectic field was represented by a single value of the phase $\Phi$ at each point on a 41 or 128 unit cubic grid with periodic boundary conditions. Discretized energy expressions were used to calculate the local values of compression, curvature, and saddle-splay energy. We manually put line defects into our lattice which forced the P~surface or I-Wp~surface topology, similar to the analysis in \cite{Holyst}. The energy was then minimized by the conjugate gradient method. Because of the troublesome nature of defining and allowing defects in such a phase field simulation, as well as the relatively small grid size, the numerical values in these simulations cannot be considered precise. However, study of the numerical minimal energy configurations are instructive for both validating and expanding upon our models. 

\begin{figure}

\epsfig{file= 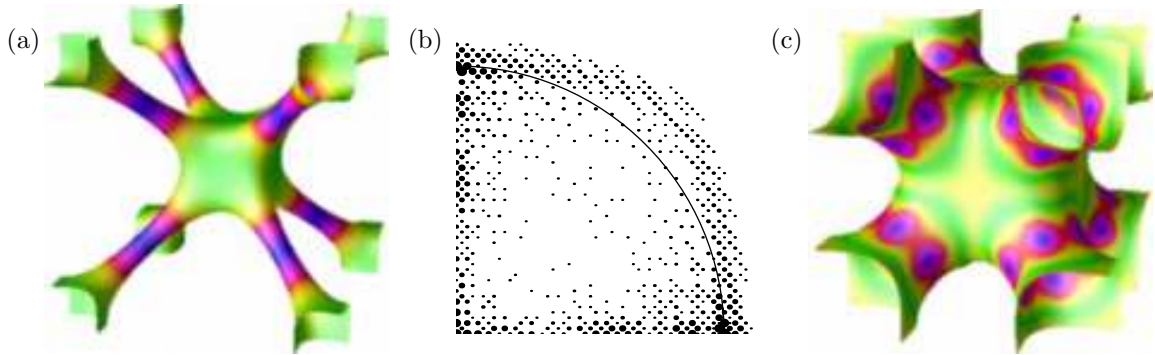}

\caption{Numerical minimization results with parallel defect boundary conditions. Images (a) and (c) show two different smectic layers for a relaxed configuration of the I-Wp surface. Surface coloring is proportional to local curvature energy density, with blue and violet denoting the highest values. Plot (b) shows the strongest Fourier components in the $k_z=0$ plane for this smectic configuration.}
\label{fig:smallcoresim}
\end{figure}

We imposed parallel boundary conditions on the smectic field at the defect lines (in other words, the defect lines become surfaces of constant phase). This corresponds to our ``small core" construction. The results are pictured in Fig.~\ref{fig:smallcoresim}. The shading indicates local bending energy density. For $B/K_1$ relatively small, the middle surface between inner and outer defect lattices is close to a minimal surface. For layers away from the middle surface, such as that pictured in  Fig.~\ref{fig:smallcoresim}(a), the curvature energy is concentrated around the cylindrical regions, while the remaining areas are close to minimal surfaces. Fig.~\ref{fig:smallcoresim}(b) shows the strong Fourier components in the $k_z=0$ plane for the I-Wp surface smectic. The numerically relaxed configurations for moderate values of $\lambda \equiv \sqrt{K_1/B} \approx 1$ and $L/a_0 \approx 40$ layers per unit cell showed the same off-peak increase for $\vert k_{max} ( \theta, \phi ) \vert$ as was found experimentally by Pansu et al.~\cite{pansubp2}.

\begin{figure}
\center

\epsfig{file= 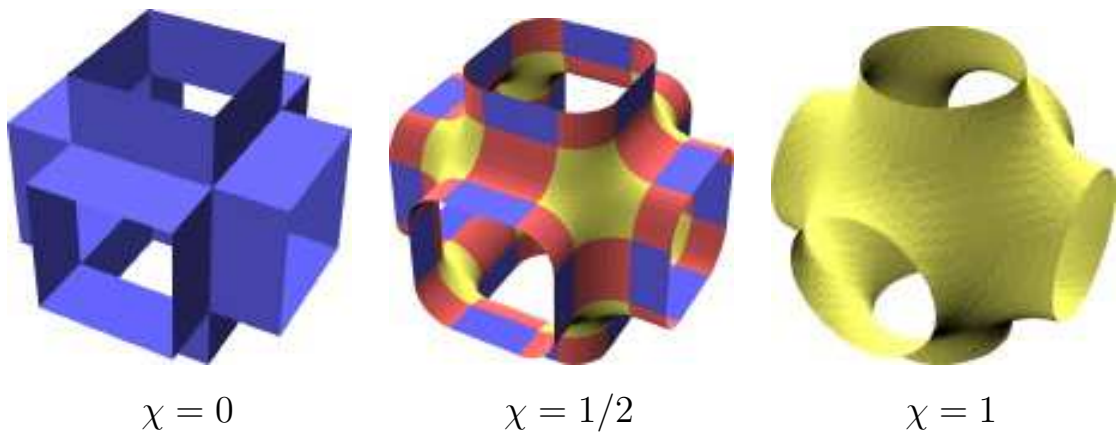}

\caption{A one parameter model for curvature condensation to balance curvature and compression energies.}
\label{fig:curvcond}
\end{figure}

For higher values of $B/K_1$, such as in Fig.~\ref{fig:smallcoresim}(c), the smoothly curved regions of our construction start to become faceted, with the curvature condensing onto discrete folds. This is not unlike the condensation of curvature found in crumpled elastic sheets~\cite{crumpling}. Motivated by this apparent tradeoff between curvature and compression energies, we examined the model for relaxation pictured in Fig.~\ref{fig:curvcond}, in which the regions of high curvature become more concentrated in space while the remaining overall surface becomes more polyhedral. This model has one continuous parameter $\chi\in[0,1]$ which interpolates between the minimal surface model at $\chi=1$ and a faceted surface with uniform layer spacing away from tilt grain boundaries at $\chi=0$. In this model, we shrink patches of the minimal surface onto vertices of a polyhedron with the same symmetry. The patches are connected by cylindrical sections, and the remaining area is filled with flat facets. An expression for the compression energy of this structure, using our earlier techniques, is complicated, but numerical evaluation for the P~surface found it was well approximated by a simple linear function of $\chi$. The curvature energy is found by adding up the contributions of the cylindrical regions. We have ignored the complicated non-linear effects at low $\chi$, so our analysis is valid near $\chi=1$. The resulting combined compression and curvature energy density for the P~surface smectic is 
\begin{equation}
F_B+F_C = 4.32 \times 10^{-2} \chi BL^3 + 6 \pi K_1 L \left[\frac{1}{\chi} \log (L/4\rho_c) + \left(4\rho_c/L-1\right)\right].
\end{equation}
Minimizing in $\chi$ gives
\begin{equation}
\chi = \sqrt{\frac{12 \pi K_1\log (L/4\rho_c)}{4.32 \times 10^{-2} BL^2}} = 0.94 \frac{\lambda}{a_0},
\label{eq:chival}
\end{equation}
where the last equality is for $L=50a_0$. Since $\lambda \ge a_0$ and this treatment is only valid for $\chi \le 1$, we see that  curvature condensation will not happen for our small smectic repeat unit, but might occur for $L\gg 50a_0$.

\section{Large core model: Calculation of curvature energies using Weierstra\ss \ representation}
\label{sec:largecore}
\subsection{Geometric Construction}
\begin{figure}
\center

\epsfig{file=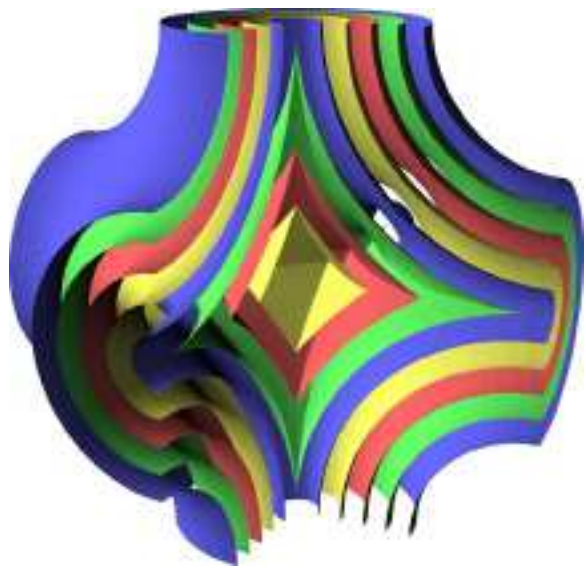, width=3in}

\caption{Construction of smectic blue phase based on uniform layer spacing of the Schwartz P~surface. The image shows a cutaway of the layered structure inside the P~surface.}
\label{fig:evenconstruction}
\end{figure}

The large core variant of our construction begins with a single continuous surface, then adds successive layers which are equally spaced along the original surface normals. As we showed in Section~\ref{sec:energetics}, the continuation of uniform layer spacing will eventually lead to singularities. Fig.~\ref{fig:evenconstruction} shows the results of this construction for the interior of a P~surface section. The initial curvature singularities occurs on the nearly circular interface between unit cells of the P~surface: essentially, the circular interface shrinks to a point, pinching off the connection between the interior layers and the outside of the unit cell. Any further layers within the pinched layer must be topologically like a sphere, with genus $0$.  Note that this layering scheme appears similar to the construction of constant mean curvature (CMC) surfaces~\cite{scriven} but is essentially different;  there is no energetic preference for constant mean curvature and the CMC surfaces do not minimize the
compression energy. We shall label the normal distance from the initial minimal surface to the last fully connected surface as $a_{max}$. Since the principal radii of curvature are equal and opposite for points on the minimal surface, the value of $\vert a_{max} \vert$ will be the same on both sides of the minimal surface. We call the region within the pinch-off surface, at distance greater than $a_{max}$ from the minimal surface,  the core region since our initial ordering is disrupted in this volume. Outside the core, the saddle-splay energy per unit cell is $F_{SS} = 16 \vert K_{24} \vert (g-1) a_{max}$. The values of $g$ and $a_{max}$ for the P~and~I-Wp~surfaces are given in Table~\ref{table:weierstrass}. Fig.~\ref{fig:skeleton}(a), (b) and (c) show the surfaces defined by $a_{max}$ for both minimal surfaces.  Within the pictured surfaces, the continuation of our initial construction produces additional curvature singularities and lines of surface self-intersection. At the end of this section we justify a new model for the core, which fills it with even layers joined by tilt grain boundaries.

By insisting on uniform spacing away from the fiducial minimal surface, we determine the exact shape of all the layers outside the core with no free parameters.
Clearly, the shape equations of all the layers can be obtained from that of the original surface if we know how all quantities evolve with normal displacement. The curvature evolution was derived in Section~\ref{sec:energetics}.
We can also map the area element of the initial surface to that on successive surfaces by $dA' = dA\times(1+2aH+a^2K_G)$. This allows us to express the curvature component of the smectic free energy (\ref{eq:smfe}) as 
\begin{equation}
F_B = 4 K_1 \int_0^{a_{max}} da \int dA  \frac{(H+a K_G)^2}{1+2aH+a^2K_G},
\label{eq:cfe}
\end{equation}
where the area integral is taken over the center surface. In our case the area integral is over the minimal surface  repeat cell and $H=0$. We evaluated the surface integral in Eq.~(\ref{eq:cfe}) analytically using the Weierstra\ss \ representation for the minimal surfaces~\cite{nitsche,pansums}, as described in Appendix~\ref{sec:weierstrass}. Table~\ref{table:weierstrass} gives numerical values for the bending energy using Eq.~(\ref{eq:cfe}) and the saddle-splay energies using the Gauss-Bonnet theorem and Eq.~(\ref{eq:ssad}). In both cases the integration limit $a_{max}$ corresponds to the value of $a$ for which $H'$ in Eq.~(\ref{eq:curvvsa}) first diverges at some point on the surface. Thus $a_{max} = \left(- K^{min}_G\right)^{-1/2}$, with $K^{min}_G$ the minimum (most negative) value of $K_G$ on the minimal surface. 

\begin{table}
\center

\renewcommand{\arraystretch}{1.5}

\begin{tabular}{| c | c | c | c | c | c | c |}\hline
surface & $\rgen$ & $a_{max}$ (L) & $F_B \ (K_1 L)$ & $F_{SS} \ (\vert K_{24} \vert L)$ & 
 $F_{GB} \ (K_1 a_0^{-1} L^2)$\\ \hline 
P & $\left(1+14\omega^4+\omega^8\right)^{-1/2}$ & $0.232$ & 12.2 & -23.3 & 1.65 \\
I-Wp &  $\left( \omega^6 - 5 \omega^4 - 5 \omega^2 +1 \right)^{-2/3}$ & 0.150 & 14.7 & -45.2 & 1.36\\ \hline 
\end{tabular}

\caption{Surface geometries and energies in the "large core" construction for the P and I-Wp surfaces. The first column gives the generating function for the Weierstra\ss \ representation, followed by the maximum half-thickness of continuous layering, the curvature energy, saddle-splay energy, and grain boundary core energy for this {\sl ansatz}.}
\label{table:weierstrass}
\end{table}

All that remains is to find the optimal way to fill the `` core" volumes for $\vert a \vert > \vert a_{max} \vert$. The shape of these regions is shown in Fig.~\ref{fig:skeleton}(a), (b), and (c). In the core, the continuation of uniform layer spacing from the minimal surface leads to curvature singularities and layer self intersections. However, as shown in Fig.~\ref{fig:skeleton}(a), we can fill this volume with uniformly spaced domains of smectic order which intersect each other in tilt grain boundaries. Within a domain, each point on a given layer is an equal normal distance away from some point on the initial minimal surface. Points on the grain boundaries between domains are an equal normal distance away from at least two different points on the minimal surface, so the grain boundaries occur where the uniform spacing used outside the core would cause layer self-intersection.  We have no general principle of why the singularities in the core region are domain walls instead of, say, focal conic domains (such as you may find by developing a catenoid). For both the P and I-Wp surfaces uniform spacing leads to disclination walls, as it probably does for all but highly symmetric surfaces. 

Thus, with the aim of avoiding layer compression, we choose this solution for the core structure, which should serve as an energetic upper bound.
The energy of the core will arise from the regular curvature and saddle-splay energies within the smectic domains, along with a surface energy at the grain boundaries and a separate line energy in the highly distorted regions where the grain boundaries meet. A conservation estimate of the energy of a tilt grain boundary is approximately $K_1 a_0^{-1}$ per unit area~\cite{kleman} (assuming a melted wall of thickness $a_0$ at the defect plane). 
Numerically, the total area of grain boundary walls per unit cell in this construction is  $A= 1.65 L^2$ for the P~surface and $A= 1.36 L^2$ for the I-Wp~surface; the corresponding energies are given in Table~\ref{table:weierstrass}. The lines at the intersection of grain boundaries are located where the $+1$ and $+1/2$ disclinations were in the small core model of Section~\ref{sec:smallcore}. Around these lines the director field varies so rapidly that the smectic order should melt completely, leaving a core of radius $\sim a_0$ and line tension $\varepsilon$. The mean curvature energy in the smectic domains is relatively small, since the shape of the innermost layers is mapped by Eq.~(\ref{eq:curvvsa}) from the region around the umbilics (flat points) on the minimal surface. The inner layers are thus nearly flat. We found that the curvature energy was negligible by numerical evaluation.

The saddle-splay energy in the core merits special attention. Since the smectic layers in the core are not closed surfaces, our simple topological arguments do not apply. We could insist on connecting the surface sections together in a natural way across grain boundaries to obtain closed surfaces with the topology of a sphere. However, these closed surfaces are nearly polygonal, with all the Gaussian curvature concentrated at the vertices, or along the lines in three dimensions where the smectic order has melted. Thus. it is natural to assume that the smectic order parameter vanishes in these line-like cores, and the total saddle-splay of the core is nearly zero.

The dominant free energy terms which determine the stability and preferred size of this construction are therefore the bend, saddle-splay and grain-boundary energies. Minimizing these energies in a unit cell of length $L$ gives:
\begin{equation}
L = (\vert F_{SS} \vert - F_B)/\left(2 F_{GB}\right). 
\label{eq:approxtotal}
\end{equation}
At this length the phase is stable against the standard, flat smectic when $\vert F_{SS} \vert > F_B $. Reading values from Table~\ref{table:weierstrass}, the minimum for the P surface structure occurs for $L = 7.06 a_0 (\vert K_{24} \vert - 0.523 K_1) / K_1 $, while that for the I-Wp occurs at 
$L = 16.6  a_0 (\vert K_{24} \vert  - 0.325 K_1) / K_1$. To compare with experiment we set $L = 50 a_0$, where $a_0$ is the smectic layer spacing. This requires $ \vert K_{24} \vert / K_1 = 6.6$ for the P surface or $2.6$ for the I-Wp surface.  Thus the I-Wp {\sl ansatz} should be the stable phase at any value of $K_{24}$.

\subsection{Numerical Minimization}
\begin{figure}

\epsfig{file= 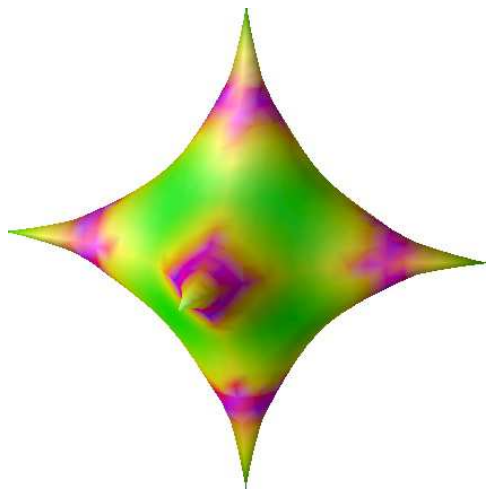, width=2.5in}\hspace{1in}

\caption{Numerical minimization results with arbitrary defect boundary conditions. The image shows the core region of a P-surface smectic.}
\label{fig:largecoresim}
\end{figure}

As before, we used conjugate-gradient minimization to explore the possible smectic configurations.  In this case we relaxed the parallel boundary conditions, letting the interface of smectic field and defect line be arbitrary. The results are shown in Fig.~\ref{fig:largecoresim}. The core region resembles our "large core" model, except that the defect walls have disappeared at the expense of greater overall compression. Locally, the core layers look like focal conics around the defect lines. This result undoubtedly points to configurations close to our model but with lower energy, and most likely stable for even lower values of $\vert K_{24}\vert /K_1$. Due to numerical difficulties, these phases could only be made stable for large values of $L/a_0 > 80$, so we could not compare configurations in the region of interest.

\section{Comparison to Spectroscopic Data}
\label{sec:xray}

\begin{figure}

\epsfig{file= 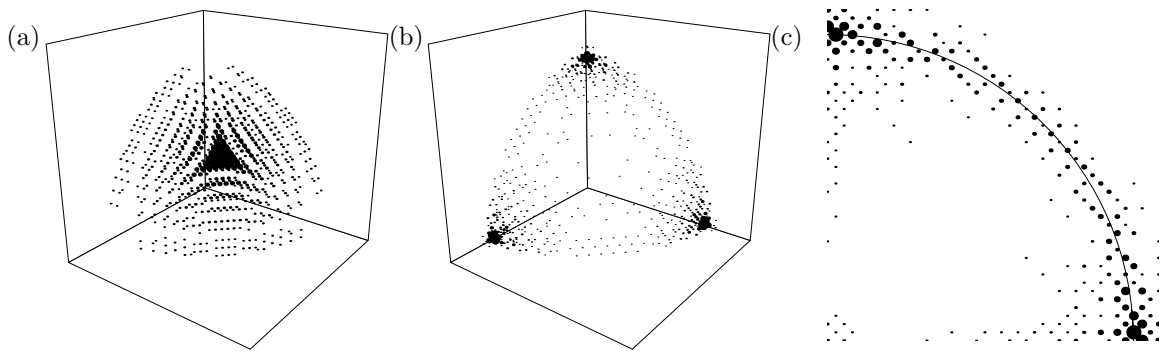}

\caption{Discrete Fourier transform of proposed smectic structures. Plot (a) is for the P surfaced based construction, (b) and (c) are for the I-Wp construction. Plots (a) and (b) show the largest Fourier components in the positive coordinate octant of $k$ space. Plot (c) shows the largest Fourier components in the positive coordinate quadrant of the $k_z=0$ plane. Point size is proportional to the Fourier coefficient at that $\vec{k}$.  }
\label{fig:fft}
\end{figure}

Pansu {\it et.al.}~\cite{pansubp1,pansubp2,pansubp3} found through X-ray diffraction that the cubic $BP_{SmA}1$ phase showed the strongest diffraction along three mutually orthogonal axes. They also found that the peak wavelength of diffraction decreased away from these special directions. In order to compare our model to their data, we have calculated the expected scattering from a smectic blue phase built by our ``large core" construction. Fig.~\ref{fig:fft} shows the $k$-space location of the strongest Fourier peaks for both a P surface smectic and an I-Wp surface smectic. In each case, we sampled the smectic phase on a cubic grid with $128$ points per side. We took the smectic phase to be $2 \pi \times 37 \times x_n / L$, where $x_n$ is the distance of the sample point normal to the minimal surface. We chose this number of layers to avoid any commensuration effects.  Because there are many layers, this
was numerically challenging and a higher precision study is needed, such as the study of scattering from triply-periodic minimal surfaces by Garstecki and Ho\l yst~\cite{holyst2}.

For both structures, and we expect for general minimal surface smectics, the Fourier peaks were strongest along the umbilics of the minimal surfaces, {\sl i.e.} normal to the points at which the surfaces are flat. The Fourier peaks of the I-Wp structure are a clear match to the observed X-ray diffraction of $BP_{SmA}1$. However, as Fig.~\ref{fig:fft}(c) shows, there is no clear deviation in diffraction maximum wavelength as a function of angle (Pansu {\it et.al.} reported a $7\%$ deviation over $25^\circ$ angle). This indicates the need for further refinement to our model if it is to explain the current data exactly. The discrepancy in off-peak behavior between our simulations and real data could arise from some relaxation of our proposed structure to better accommodate the competing energies. As we noted in Section~\ref{sec:simulation}, the numerical relaxation of the I-Wp smectic with parallel (``small core'') boundary conditions does produce the correct off-peak behavior for the diffraction maxima.

\section{Conclusion}

We have presented a new model for smectic blue phases which is stable for physically realizable values of $K_{24}$ and penetration length $\lambda \equiv \sqrt{K_1/B}$. This work refines that of our previous paper~\cite{didonna} and gives further weight to our proposed organizing principle of smectics built on minimal surfaces.  Unlike the traditional blue phases, our model does not rely on molecular
chirality.  It would be interesting to add chirality into this model to see how the new length
scale alters the equilibrium structures.
We have also addressed the broader question of geometrical frustration between Gaussian curvature and smectic order.  Our computed structure factor matches the scattering from the $Sm_{BP}1$ phase and so we remain optimistic that our construct could be verified, perhaps
through freeze-fracture.  

\acknowledgments
We thank T.C. Lubensky,  B. Pansu and J.P. Sethna for fruitful discussions.  This work
was supported by NSF Grants INT99-10017 and DMR01-29804, the Donors of the Petroleum Research Fund Administered by the American Chemical Society, the Pennsylvania Nanotechnology Institute, and a gift from L.J. Bernstein.

\appendix

\section{Alternate derivation of curvature evolution: Layering and curvature frustration in curved space}

\begin{figure}

\epsfig{file= 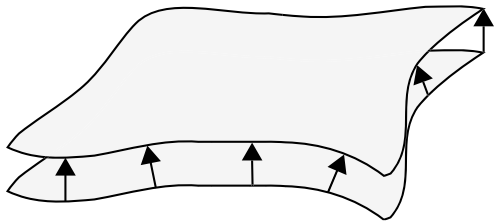, width=2.5in}\hspace{1in}

\caption{One patch being uniformly translated along its normal.  The top and bottom surfaces along
with the perpendicular surfaces generated by the normals at the boundaries makes an integration
volume $V$.}
\label{gaussshift}
\end{figure}

The first result in Eq.~(\ref{eq:deltacurv}) also follows by considering the derivative of the mean curvature along the layer normal:
\begin{equation}
\label{other}
\frac{\partial H}{\partial a}=\left({\bf N}\cdot\nabb\right) \left[\frac{1}{2}\nabb\cdot{\bf N}\right] = \frac{1}{2}\nabb\cdot\left[ {\bf N}\left(\nabb\cdot{\bf N}\right)\right] -\frac{1}{2}\left(\nabb\cdot{\bf N}\right)^2
\end{equation}
When the layers are built by developing surfaces parallel to a fiducial surface, the layer
normals do not change -- in other words, since we are translating the surface parallel to ${\bf N}$, we must have $\left({\bf N}\cdot\nabb\right){\bf N}=0$~\cite{sethnakleman}.  Adding this term to Eq.~(\ref{other})
we have:
\begin{eqnarray}
\label{otherlater}
\frac{\partial H}{\partial a}&=&  \frac{1}{2}\nabb\cdot\left[ {\bf N}\left(\nabb\cdot{\bf N}\right)-\left({\bf N}\cdot\nabb\right){\bf N}\right] -\frac{1}{2}\left(\nabb\cdot{\bf N}\right)^2\nonumber\\
&=& K_G - 2H^2
\end{eqnarray}
Similarly, consider a patch on one layer and the corresponding patch on a uniformly
translated surface as shown in Fig.~\ref{gaussshift}.  Because the principle directions of the two surfaces are 
unchanged, the geodesic curvature of the corresponding boundaries are identical.  It
follows from the Gauss-Bonnet theorem that $\int_{M_{1,2}} K_G dA$ is the same on the
two patches.  Thus, over the volume $V$ swept out by making the translated patch (shown in Fig.~\ref{gaussshift}), $\int_V  K_G {\bf N}\cdot {\bf dA} =0$.  Therefore $\nabb\cdot \left(K_G{\bf N}\right)=0$
and
\begin{eqnarray}
\label{lastlater}
0&=& {\bf N}\cdot\nabb K_G + K_G\nabb\cdot{\bf N}\nonumber\\
0&=&\frac{\partial K_G}{\partial a} + 2K_G H
\end{eqnarray}

The intrinsic frustration between curvature and uniform layer spacing can be relieved by considering layered systems in curved space, just as double-twist in the classical blue phases can fill the surface of the three-dimensional sphere without defects~\cite{sethna-curved}.   It is straightforward to see this using the derivations above.  In curved space, we simply replace derivatives  with {\sl covariant} derivatives, so that $\nabb\rightarrow {\bf D}$ where ${\bf D}$ is 
the covariant derivative~\cite{misner}.  The mean curvature is $H={\frac{1}{2}}D_iN^i$ and is essentially unchanged.  However, the relation between Gaussian curvature and saddle-splay is more subtle.  Recall that in flat space, the coefficients of the two terms in the saddle-splay are
constrained so that the saddle-splay only depends on first derivatives of ${\bf N}$.  We shall see that this is spoiled in curved space.  We have:
\begin{eqnarray}
{\bf D}\cdot\left[ \left( {\bf N} \cdot {\bf D} \right) {\bf N} - {\bf N} \left( {\bf D}\cdot {\bf N} \right) \right]&=&
D_i\left[N^jD_jN^i - N^iD_jN^j\right]\nonumber\\
&=&\left( D_iN^jD_jN^i-D_iN^iD_jN^j\right) + N^j\left[D_i,D_j\right]N_i
\label{thisisit}
\end{eqnarray}
The first term in~(\ref{thisisit}) is $-2K_G$, a factor times the Gaussian curvature.  Note that in
flat space, the covariant derivatives become simple derivatives and commute.  In curved space
this is no longer true and the commutator term is just $R_{NN}$, the component of the Ricci tensor in the normal-normal direction.   The evolution equation for $K_G$ is unchanged since Stoke's theorem (properly modified) holds in curved space.  Thus in curved space we have
\begin{eqnarray}
\frac{\partial K_G}{\partial a} &=& -2K_GH\nonumber\\
\frac{\partial H}{\partial a}&=& K_G - 2H^2 -{\frac{1}{2}} R_{NN}
\end{eqnarray}
Thus we can have uniformly spaced layers with $H=0$ and $K_G$ constant ($K_G$ may vary in the layer, but will not change from layer to layer) if 
we embed the smectic in a space with $R_{NN}=2K_G$.  Since we are considering surfaces with $K_G<0$, this suggests a space with negative curvature.  

To see this, we consider a special coordinate system in a $3$-dimensional layered structure. For unbroken layers we can uniquely define the continuous variable $\Phi$ which labels the layers. Furthermore, we can use $\Phi$ as a local coordinate to define a coordinate basis in which one basis vector $e_3$ is dual to the directional derivative in $\Phi$, and the other basis vectors $e_1$ and $e_2$ lie in planes of constant $\Phi$. This coordinate system is known as Gaussian normal coordinates~\cite{misner}. Furthermore, for uniform layer spacing the measurement of distance along $e_3$ cannot depend on coordinates $1$ and $2$. This, plus the orthogonality of  $e_3$ lets us scale our coordinates such that $g_{3 i} = e_3 \cdot e_i = \delta_{3i}$. 

Since we are using a coordinate basis, the covariant derivative is defined with the typical connection coefficients $\Gamma^i_{\ jk}$. Elementary considerations relate the connection coefficients to the extrinsic curvature tensors of the layers taken as $2$-sheets:
\begin{gather}
\kappa_{ij} = \Gamma_{3ij} = - \Gamma_{i3j}  = - \Gamma_{j3i} \textrm{ for } i,j \ne 3 \nonumber \\
\Gamma_{i33}=\Gamma_{3i3}=\Gamma_{33i}=0.
\end{gather}
Laborious but autonomic calculations show that the Ricci scalar $R=2R_{NN}$ and so
the space has negative scalar curvature.  This is not surprising: in the classical blue phases
the saddle-splay elastic constant needed to be {\sl positive}, favoring positive Gaussian
curvature, and the resulting structure could be defect-free in positively curved space.  We have
merely ``flipped the signs'' on the last sentence.  Further work on flattening the hyperbolic space
into ${\bf R}^3$ would be interesting and may shed light on the preferred lattices.

\section{Weiersta\ss \ representation of minimal surface}
\label{sec:weierstrass}

\begin{figure}
\center

\epsfig{file=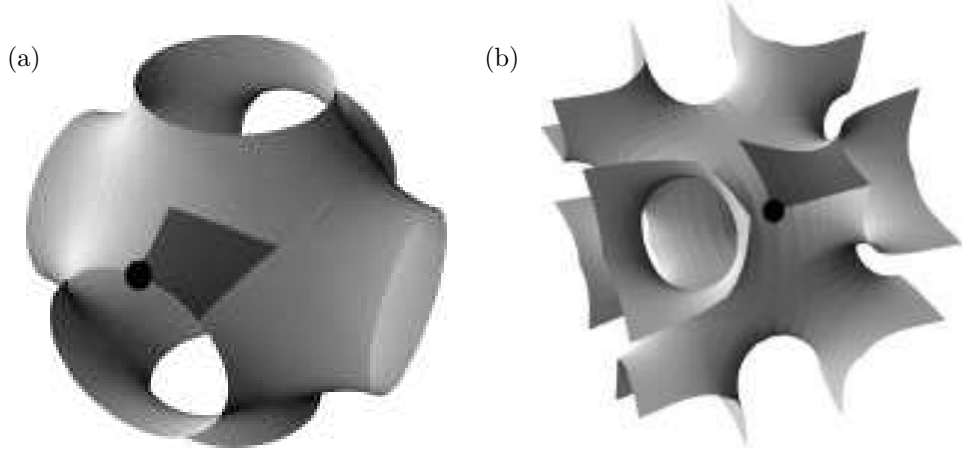}

\caption{Weierstra\ss \ representation of the P~surface, (a), and I-Wp~surface, (b). The shaded region in each is 1/48 part of the pictured surface repeat unit and contains all the symmetry of the entire surface. The points represent the location of $\omega = 0$ for the Weierstra\ss \ mappings given in Table~\ref{table:weierstrass}}
\label{fig:weirp}
\end{figure}

The Weierstra\ss \ representation of minimal surfaces relies on the fact that there is locally a one to one mapping from a minimal surface onto the unit sphere via the surface normal \cite{nitsche}. The unit sphere can, in turn, be mapped onto the complex plane by stereographic projection, with the plane passing through the equator of the sphere. The latter mapping gives the components of the surface normal in terms of the complex variable $\omega = \sigma + i \tau$ as 
\begin{equation}
N_1  =  \frac{2 \Re \ \omega}{1+\vert \omega \vert^2}, \hspace{.5in}
N_2  =  \frac{2 \Im \ \omega}{1+\vert \omega \vert^2}, \hspace{.5in}
N_3  =  \frac{1-\vert \omega \vert^2}{1+\vert \omega \vert^2}
\end{equation}
The reverse mapping between the complex plane and the minimal surface itself can be expressed via a single generating function $\rgen$.  In terms of $\rgen$, displacements on the minimal surface are
\begin{gather}
dx^1  =  \Re \left[\left(1-\omega^2\right) \rgen d\omega \right] \nonumber \\
dx^2 =  -\Im \left[\left(1+\omega^2\right) \rgen d\omega \right] \nonumber \\
dx^3  =  -2\Re \left[ \omega  \rgen d\omega \right] 
\label{eq:rgenmap}
\end{gather}
The relative coordinates of points on the surface can be found by integrating Eq.~(\ref{eq:rgenmap}). 

To complete our reparameterization of Eq.~(\ref{eq:cfe}) in terms of $\omega$, we quote expressions for other quantities of interest on the surface~\cite{nitsche,pansums}:
\begin{gather}
dA = \vert \rgen \vert^2 \left(1+ \vert \omega \vert^2 \right)^2 d\sigma d\tau  \nonumber \\
K_G = -4 \vert \rgen \vert^{-2}\left(1+ \vert \omega \vert^2 \right)^{-4}
\label{eq:rgenshape}
\end{gather}

The Weierstra\ss \ representations for the surfaces we consider are given in Table~\ref{table:weierstrass}. For the P~surface, $\omega=0$ ($z=-1$ on the unit sphere) corresponds to the point indicated in Fig.~\ref{fig:weirp}(a). The entire shaded region in Fig.~\ref{fig:weirp}(a), which is $1/48$ of the P~surface repeat cell, maps to the region on the unit sphere bounded by the intersection of the sphere with the planes $x=0$, $y=0$, $z=-x$ and $z=-y$. For the I-Wp surface, the position of $\omega=0$ is shown in Fig.~\ref{fig:weirp}(b). The shaded region is again $1/48$ of the I-Wp repeat cell and corresponds to the region on the unit sphere bounded by the planes $x=0$, $y=0$, $z=x$, and $z=\sqrt{2} y - x$.


\begin{thebibliography}{99}


\bibitem{spaghetti1}
Z. Kutnjak, C. W. Garland, J. L. Passmore, and P. J. Collings, 
\prl {\bf 74}, 4859 (1995); J. B. Becker and P. J. Collings, Mol. Cryst. Liq. Cryst. {\bf 265}, 163 (1995).


\bibitem{spaghetti2}
H.-S. Kitzerow and P.P. Crooker, \prl {\bf 67}, 2151 (1991); 
H.M. Hornreich, \prl {\bf 67}, 2155 (1991).

\bibitem{mhli} 
B. Pansu, M.-H. Li,  and H.T. Nguyen,  
J. Phys. II France {\bf 7}, 751 (1997).

\bibitem{grelet}
E. Grelet, B. Pansu, M.-H. Li and H.T. Nguyen, \pre {\bf 65} 050701(R) (2002).

\bibitem{pansubp1} 
E. Grelet, B. Pansu, M.-H. Li, and H.T. Nguyen, 
\prl {\bf 86}, 3791 (2001).

\bibitem{pansubp2}
B. Pansu, E. Grelet, M.-H. Li, and H.T. Nguyen, 
\pre {\bf 62}, 658 (2000).

\bibitem{pansubp3} 
E. Grelet, B. Pansu, and H.T. Nguyen, 
\pre {\bf 64}, 010703(R) (2001).

\bibitem{kamien} 
R.D. Kamien, J. Phys. II France {\bf 7}, 743 (1997).
	
\bibitem{sethna} 
S. Meiboom, 
J.P. Sethna, P.W. Anderson, and W.F. Brinkman, 
\prl {\bf 46}, 1216 (1981).

\bibitem{didonna} B.A. DiDonna and R.D. Kamien, \prl {\bf 89}, 215504 (2002).

\bibitem{rmp}
R.D. Kamien, \rmp {\bf 74}, 953 (2002).

\bibitem{foot1} When $g=0$ the unit cell must have the topology of a sphere and cannot have any handles or holes to allow it to connect to the neighboring unit cells.

\bibitem{guven}
R. Capovilla and J. Guven J Phys A {\bf 35}, 6233 (2002).

\bibitem{visuals} The Scientific Graphics Project, "http://www.msri.org/publications/sgp/SGP/indexc.html"

\bibitem{Brakke92} 
K. Brakke, Exp. Math. {\bf 1}, 141 (1992).

\bibitem{Holyst} W. G\'ozdz and R. Ho\l yst, \pre {\bf 54}, 5012 (1996).

\bibitem{crumpling} B.A. DiDonna, \pre {\bf 66}, 016601 (2002). 

\bibitem{scriven}
D. Anderson, H. Davis, L. Scriven, and J.C.C. Nitsche,
Adv. Chem. Phys. {\bf 77}, 337 (1990).

\bibitem{nitsche}
J.C.C. Nitsche, {\sl Vorlesungen \"uber Minimalfl\"achen},
Springer-Verlag, Berlin, (1975); J.C.C. Nitsche, {\sl Lectures on Minimal
Surfaces},
(Translated by J.M.~Feinberg), Cambridge University Press, Cambridge, (1989).

\bibitem{pansums} 
B. Pansu and E. Dubois-Violette,  Europhys Lett. {\bf 10}, 43 (1989).

\bibitem{kleman} M. Kl\'eman, {\it Points, Lines  and Walls}, John Wiley and Sons, New York (1983).

\bibitem{holyst2} P. Garstecki and R. Ho\l yst, \pre {\bf 64}, 021501 (2001).

\bibitem{sethnakleman} J. P. Sethna and M. Kl\'eman, \pra {\bf 26}, 3037.
	
\bibitem{sethna-curved} J. P. Sethna, D. C. Wright, and N. D. Mermin, \prl {\bf 51}, 467 (1983).


\bibitem{misner} C. W. Misner, K. S. Thorne, and J. A. Wheeler, {\it Gravitation} W. H. Freeman and Co., New York  (1973).



\end{thebibliography}
\end{document}